\pdfoutput=1
\documentclass[11pt,twoside,a4paper,cmspaper,final,collab]{cms-tdr}
\def\svnVersion{e10bba6}\def\svnDate{2024/08/10}\def\cmsCernNoTag{CERN-EP-2024-006}\def\cmsCernDate{\today}\def\cmsMessage{Published in the European Physical Journal C as \href{http://dx.doi.org/10.1140/epjc/s10052-024-13114-9}{\doi{10.1140/epjc/s10052-024-13114-9}.}}
\begin{document}\cmsNoteHeader{BPH-22-002}

\DeclareRobustCommand{\Xibarp}    {{\HepParticle{\PAGX}{}{+}}\Xspace}

\newcommand{\mupm}      {\PGmpm}
\newcommand{\mumu}      {\ensuremath{\PGmp\PGmm}\xspace}
\newcommand{\pip}       {\ensuremath{\PGpp}\xspace}
\newcommand{\pim}       {\ensuremath{\PGpm}\xspace}
\newcommand{\pipi}      {\ensuremath{\pip\pim}\xspace}
\newcommand{\Kp}        {\PKp}
\newcommand{\Km}        {\PKm}
\newcommand{\Bm}        {\ensuremath{\PBm}\xspace}
\newcommand{\Bs}        {\ensuremath{\PBzs}\xspace}
\newcommand{\Lb}        {\ensuremath{\PGLzb}\xspace}
\newcommand{\Xibm}      {\ensuremath{\PGXmb}\xspace}
\newcommand{\prot}      {\ensuremath{\Pp}\xspace}
\newcommand{\aprot}     {\ensuremath{\PAp}\xspace}
\newcommand{\La}        {\ensuremath{\PGL}\xspace}
\newcommand{\Xim}       {\ensuremath{\PGXm}\xspace}
\newcommand{\Omm}       {\ensuremath{\PGOm}\xspace}
\newcommand{\jpsi}      {\ensuremath{\PJGy}\xspace}
\newcommand{\psit}      {\ensuremath{\PGyP{2\cmsSymbolFace{S}}}\xspace}
\newcommand{\JMM}       {\ensuremath{\jpsi\to\mumu}\xspace}
\newcommand{\PJPP}      {\ensuremath{\psit\to\jpsi\pipi}\xspace}
\newcommand{\JPP}       {\ensuremath{\jpsi\pipi}\xspace}
\newcommand{\mJPP}      {\ensuremath{m(\JPP)}\xspace}
\newcommand{\pp}        {\ensuremath{\prot\prot}\xspace}
\newcommand{\XiLaPi}    {\ensuremath{\Xim\to\La\pim}\xspace}
\newcommand{\LaPPi}     {\ensuremath{\La\to\Pp\pim}\xspace}
\newcommand{\LbPsiLam}  {\ensuremath{\Lb\to\psit\PGL}\xspace}
\newcommand{\LbJpPrK}   {\ensuremath{\Lb\to\jpsi\Pp\Km}\xspace}
\newcommand{\LbJpXiK}   {\ensuremath{\Lb\to\jpsi\Xim\Kp}\xspace}
\newcommand{\PsiLam}    {\ensuremath{\psit\PGL}\xspace}
\newcommand{\JpXiK}     {\ensuremath{\jpsi\Xim\Kp}\xspace}
\newcommand{\rr}        {\ensuremath{\mathcal{R}}\xspace}
\newcommand{\chisq}     {\ensuremath{\chi^2}\xspace}
\newcommand{\pvtx}      {\ensuremath{P_{\text{vtx}}}\xspace}
\newcommand{\tev}       {\TeV}
\newcommand{\mev}       {\MeV}
\newcommand{\ptmu}      {\ensuremath{\pt(\mupm)}\xspace}
\newcommand{\ptk}       {\ensuremath{\pt(\Kp)}\xspace}
\newcommand{\ptlb}      {\ensuremath{\pt(\Lb)}\xspace}
\newcommand{\ptjp}      {\ensuremath{\pt(\jpsi)}\xspace}
\newcommand{\ptxim}      {\ensuremath{\pt(\Xim)}\xspace}
\newcommand{\ptla}      {\ensuremath{\pt(\La)}\xspace}
\newcommand{\splot}     {\ensuremath{{}_\mathrm{{s}}\mathcal{P}\mathrm{lot}}\xspace}
\newcommand{\ebran}     {\ensuremath{(\mathcal{B})}\xspace}
\newlength\cmsTabSkip\setlength{\cmsTabSkip}{1ex}
\ifthenelse{\boolean{cms@external}}{
    \providecommand{\cmsTable}[1]{\resizebox{\columnwidth}{!}{#1}}    }
   {\providecommand{\cmsTable}[1]{#1}    }
\ifthenelse{\boolean{cms@external}}{\providecommand{\cmsLeft}{upper\xspace}}{\providecommand{\cmsLeft}{left\xspace}}
\ifthenelse{\boolean{cms@external}}{\providecommand{\cmsRight}{lower\xspace}}{\providecommand{\cmsRight}{right\xspace}} 

\cmsNoteHeader{BPH-22-002}
\title{Observation of the \texorpdfstring{\LbJpXiK}{Lambda-b0 to J/psi Xi- K+} decay}

\author*[cern]{A. Cern Person}

\date{\today}

\abstract{Using proton-proton collision data corresponding to an integrated luminosity of 140\fbinv collected by the CMS experiment at $\sqrt{s}= 13\TeV$, the \LbJpXiK decay is observed for the first time, with a statistical significance exceeding 5 standard deviations. The relative branching fraction, with respect to the \LbPsiLam decay, is measured to be $\mathcal{B}(\LbJpXiK)/\mathcal{B}(\LbPsiLam) = [3.38\pm 1.02\pm 0.61\pm0.03]\%$, where the first uncertainty is statistical, the second is systematic, and the third is related to the uncertainties in $\mathcal{B}(\PJPP)$ and $\mathcal{B}(\XiLaPi)$.
}

\hypersetup{%
pdfauthor={CMS Collaboration},%
pdftitle={Observation of the Lambda-b0 to J/psi Xi- K+ decay},%
pdfsubject={CMS},%
pdfkeywords={CMS, B hadrons, flavor physics, hadron spectroscopy, experimental results}} 

\maketitle 

\section{Introduction}

Multibody decays of beauty hadrons present a rich laboratory to search for 
intermediate resonances in the decay products. When decay products 
contain a charmonium state, such intermediate
resonances could decay into a charmonium meson and a hadron, which could be a
manifestation of their exotic nature.
An important turning point in exotic spectroscopy was achieved at the LHC, 
when the LHCb Collaboration reported the observation of statistically significant $\jpsi\prot$ 
pentaquark-like structures in the decay of the lightest beauty baryon $\LbJpPrK$~\cite{LHCb:2015yax}. 
Various interpretations of these structures have been proposed~\cite{Brambilla:2019esw,Chen:2022asf}, including tightly 
bound hidden-charm $[\PQc\PAQc\PQu\PQu\PQd]$ pentaquark states~\cite{Maiani:2015vwa,Santopinto:2016pkp}, 
loosely bound molecular baryon-meson states~\cite{Roca:2015dva,Eides:2017xnt,Du:2019pij}, 
or being due to a double triangle singularity~\cite{Nakamura:2021qvy}. 
More recently, additional exotic states have been reported by LHCb in the decays \LbJpPrK~\cite{LHCb:2019kea}, 
$\Xibm\to\jpsi\La\Km$~\cite{LHCb:2020jpq}, $\Bs\to\jpsi\prot\aprot$~\cite{LHCb:2021chn}, and 
$\Bm\to\jpsi\La\aprot$~\cite{LHCb:2022jad}.
Up to now, the hidden-charm pentaquark candidates have been reported only in $\jpsi\prot$ and $\jpsi\La$ systems. 
Investigation of other channels with heavier baryons in the decay products, such as \Xim and \Omm, could unveil 
the existence of doubly or triply strange pentaquarks~\cite{Azizi:2021pbh,Wang:2021hql}.

In this paper, we report on the search for the \LbJpXiK decay, where the \JMM, \XiLaPi, and \LaPPi channels are 
used to reconstruct the intermediate decay products. Charge-conjugate states are implied throughout the text. 
The measurement of the ratio of branching fractions
\begin{linenomath}
\begin{equation}\begin{aligned}
\ifthenelse{\boolean{cms@external}}
{ 
\rr &\equiv \frac{\mathcal{B}(\LbJpXiK)}{\mathcal{B}(\LbPsiLam)}
    = \frac{N(\LbJpXiK)}{N(\LbPsiLam)}  \\ 
    &\times\frac{\epsilon_{\PsiLam}}{\epsilon_{\JpXiK}}
    \,\frac{\mathcal{B}(\PJPP)}{\mathcal{B}(\XiLaPi)} \label{eq:R}
} 
{ 
\rr  \equiv \frac{\mathcal{B}(\LbJpXiK)}{\mathcal{B}(\LbPsiLam)}
    = \frac{N(\LbJpXiK)}{N(\LbPsiLam)}\frac{\epsilon_{\PsiLam}}{\epsilon_{\JpXiK}}
    \,\frac{\mathcal{B}(\PJPP)}{\mathcal{B}(\XiLaPi)} \label{eq:R}
} 
\end{aligned}\end{equation}
\end{linenomath}
is also reported, where $N$ is the measured \Lb yield and 
$\epsilon$ is the total efficiency. The normalization channel is chosen to be \LbPsiLam,
with the subsequent \PJPP and \JMM decays, because of its similar decay topology and 
kinematics to the signal decay, leading to the reduction of many systematic uncertainties. The 
branching fractions of the intermediate decays $\mathcal{B}(\JMM)$ and $\mathcal{B}(\LaPPi)$ 
cancel in the ratio. Invariant mass distributions of the three two-body combinations 
for the signal channel are also presented in order to look for intermediate resonances.

The analysis uses proton-proton (\pp) collision data recorded by the CMS experiment in 2016--2018, 
at $\sqrt{s}=13\tev$, corresponding to an integrated luminosity 
of $140\fbinv$~\cite{Sirunyan:2759951,CMS-PAS-LUM-17-004,CMS-PAS-LUM-18-002}. 
Tabulated results are provided in the HEPData record for this analysis~\cite{hepdata}.

\section{The CMS detector and simulated event samples}

The central feature of the CMS apparatus is a superconducting solenoid of 6\unit{m} 
internal diameter, providing a magnetic field of 3.8\unit{T}. Within the solenoid volume 
are a silicon pixel and strip tracker, a lead tungstate crystal electromagnetic 
calorimeter, and a brass and scintillator hadron calorimeter, each composed of a barrel 
and two endcap sections. Forward calorimeters extend the pseudorapidity coverage provided 
by the barrel and endcap detectors. Muons are measured in gas-ionization detectors 
embedded in the steel flux-return yoke outside the solenoid. A more detailed description 
of the CMS detector, together with a definition of the coordinate system used and the 
relevant kinematic variables, can be found in Ref.~\cite{CMS:2008xjf}.

Muons are measured in the pseudorapidity range $\abs{\eta} < 2.4$, with detection planes 
made using three technologies: drift tubes, cathode strip chambers, and resistive-plate 
chambers.
Matching muons to tracks measured in the silicon tracker results in a transverse 
momentum (\pt) resolution for muons with \pt up to 100\GeV of 1\% in the barrel and 3\% in 
the endcaps. 
The silicon tracker used in 2016 measured charged particles within the range $\abs{\eta} < 2.5$. 
For nonisolated particles of $1 < \pt < 10\GeV$ and $\abs{\eta} < 1.4$, the track 
resolutions were typically 1.5\% in \pt and 25--90
\mum in the transverse 
impact parameter~\cite{Chatrchyan:2014fea}. At the start of 2017, a new pixel 
detector was installed~\cite{Phase1Pixel}; the upgraded tracker measured nonisolated particles 
of $1 < \pt < 10\GeV$ up to $\abs{\eta} < 3$ with typical resolutions of 1.5\% in \pt and 20--75\mum in 
the transverse impact parameter~\cite{DP-2020-049}.

Events of interest are selected using a two-tiered trigger system~\cite{CMS:2016ngn}. The first level, 
composed of custom hardware processors, uses information from the calorimeters and muon 
detectors to select events at a rate of around 100\unit{kHz} within a fixed latency 
of about 4\mus~\cite{CMS:2020cmk}. The second level, known as the high-level trigger 
(HLT), consists of a farm of computing processors running a version of the full event 
reconstruction software optimized for fast processing, and reduces the event rate to 
around 1\unit{kHz} before data storage.
All events used in this analysis are selected by a set of triggers requiring two identified muons 
of opposite charge plus an additional track to form a secondary vertex, displaced from the region of 
the \pp interactions. The trigger demanded for each muon to have $\pt>4\GeV$ and 
to pass within 2\unit{cm} of the beam axis. The dimuon system was required to have 
$\pt>6.9\GeV$, invariant mass between 2.9 and 3.3\gev,
a vertex fit  probability greater than 10\%, a separation of the secondary vertex relative 
to the beam axis in the transverse plane larger than 3 standard deviations (\unit{s.d.}), and a cosine of the angle 
in the transverse plane between the dimuon momentum vector and the vector joining the beam axis and the 
dimuon vertex greater than 0.9. The additional track was required to have 
$\pt>0.8\,(1.2)\gev$ and an impact parameter with respect to the beam axis 
greater than 0 (2)\unit{s.d.}, for data collected in 2016 (2017--2018). 
Finally, the two muons and the additional track were required to originate from the same 
vertex with a \chisq per degree of freedom (dof) less than 10.

Monte Carlo (MC) simulated event samples are generated with \textsc{Pythia} 
v8.240~\cite{Pythia8} using the CP5 underlying event tune~\cite{CMS:2019csb}.
The \EVTGEN 1.6.0~\cite{Lange2001152} program models the beauty baryon decays 
\LbJpXiK and \LbPsiLam with a phase space decay model, followed by the \PJPP and \JMM decays.
Final-state radiation is included in \EVTGEN using 
\PHOTOS 3.61~\cite{Barberio:1993qi}. 
The events are then passed through a detailed \GEANTfour-based 
simulation~\cite{Agostinelli2003250} 
of the CMS detector, including also the decays of long-lived 
hyperons \XiLaPi and \LaPPi, followed by the trigger and reconstruction 
algorithms identical to those used for the collision data. 
The simulation includes additional interactions due to multiple \pp
collisions in each bunch crossing, with the same distribution as observed in the experiment.

\section{Event reconstruction and selection\label{sec:evesel}}

The reconstruction for all the decays considered in this analysis starts by finding two 
muons of opposite charge, which must match those that triggered 
the event readout and pass the soft-muon identification criteria~\cite{CMS:2018rym}.
The offline selection for both muons requires $\ptmu>3\gev$, $|\eta(\mupm)|<2.4$, 
\chisq fit probability to a common dimuon vertex $\pvtx(\mumu) > 1\%$, 
and dimuon invariant mass $2.9<m(\mumu)<3.3\gev$.

The \LaPPi candidates are selected from displaced two-prong vertices as 
described in Ref.~\cite{Khachatryan:2010pw}. 
The track with the higher momentum is assumed to be the proton one, and together with 
the pion track it is fit to a common vertex with their invariant
mass constrained to the known \La hyperon mass of $m_{\mathrm{PDG}}(\La) = 1115.683\mev$~\cite{PDG}. 
The \chisq fit probability for the \La vertex is required to be $\pvtx(\prot\pim) > 1\%$. 

{\tolerance=1600
For the signal channel, to form the \XiLaPi candidates, 
an additional high-purity~\cite{Chatrchyan:2014fea} track 
assumed to be a pion is selected with $\pt>0.2\gev$.
This track and the selected \La candidate are then fit to a common vertex 
with the $\La\pim$ mass constrained to the known \Xim hyperon mass of $m_{\mathrm{PDG}}(\Xim) = 1321.71$\mev~\cite{PDG}. 
To form the \LbJpXiK candidate, a high-purity track is chosen with an assigned kaon mass 
and $\pt(\Kp)>1.2\gev$, which aligns with the HLT \pt requirement. 
The final reconstruction step in the signal channel is the $\mumu\Xim\Kp$ vertex fit
with a \chisq probability above 1\%, where
the dimuon mass is constrained to the world-average \jpsi meson mass of $3096.9\mev$~\cite{PDG}.  
\par}

For the normalization channel, two high-purity tracks of opposite charges with 
$\pt>0.4\gev$, assumed to be pions from the \PJPP decay, are selected. 
One of them is required to have $\pt>1.2\gev$ to match the HLT \pt requirement.
The \Lb candidates are obtained by a vertex fit of the $\mumu\pipi\La$ system 
with a \JMM mass constraint, as for the signal channel.
The invariant mass of the \JPP candidates is required to be in the range 
$3.60<m(\JPP)<3.95\gev$.

From all reconstructed \pp collision points in each event, the primary vertex (PV) is chosen as the one 
with the smallest \Lb pointing angle, which is the angle between the momentum of the \Lb candidate and 
the vector from the PV to the reconstructed \Lb candidate vertex.
If any of the tracks used in the 
\Lb candidate reconstruction were included in the fit of the chosen 
PV, they are removed, and the PV is refitted.

{\tolerance=800
Selection criteria for the signal channel \LbJpXiK are optimized using the Punzi figure of 
merit~\cite{Punzi:2003bu}. The signal efficiency is evaluated using simulated event samples. 
Estimation of the background yield involves combining the collision data from the 
\Lb mass sideband, excluding the signal region which spans twice the mass resolution 
around the known \Lb mass. Additionally, the wrong-sign candidates ($\jpsi\Xim\Km$ and $\jpsi\Xibarp\Kp$) from the full mass range are included, 
after ensuring that the mass distribution of the wrong-sign candidates matches that of the correct-sign ones.
Combining these two background sources reduces the impact of the statistical uncertainty in the optimization procedure. The variables used in the 
optimization include the \pt of all decay products; the flight length 
significance in the transverse plane of the \Lb, \La, and \Xim baryon candidates and the corresponding pointing angles; 
the impact parameter significance with respect to the PV in the 
transverse plane for the tracks; 
the vertex fit probabilities; and the mass windows for hyperon candidates.
The order of the cuts is determined randomly, and in several rounds of optimization this order was different each time; 
all rounds have converged to the same final set of optimized cuts.
The resulting criteria are summarized in Table~\ref{tab:sel}.
The background is reduced by a factor of 15 after the optimization, whereas the signal efficiency is 70\% 
of the initial selection described above.
The selection criteria in the 
normalization channel \LbPsiLam are chosen to be the same, wherever possible, as in the signal channel,
to reduce the systematic uncertainties. The \JPP mass is required to be 
within 11.1\mev of the known \psit meson mass of $3686.1\mev$~\cite{PDG}, which corresponds to approximately 2.5 times the mass resolution. 
\par}

For the measurement of \rr defined in Eq.~\eqref{eq:R}, the pion from 
the \Xim decay is required to have $\pt>0.4\gev$. 
Additionally, the HLT requirements are repeated offline by requiring $\pt(\mu)>4\GeV$, $\ptjp>6.9\GeV$, 
$\pvtx(\mumu)>5\%$, and track (kaon for the signal channel, the harder of the two pions in the normalization channel) 
impact parameter above 2\unit{s.d.} with respect to the PV.
These extra criteria ensure that events from potentially inadequately modeled phase space regions are 
avoided, as the reliability of the efficiency evaluation from simulated samples in those regions is questionable.
Nevertheless, the reconstruction algorithm works reliably in those regions, and thus the corresponding events 
are used to study the mass distribution, as discussed in the following section.

In less than 5\% of the events, multiple \Lb candidates in the same channel are found. 
The rate is consistent in both channels and all candidates are used in the analysis.  
Selecting a single candidate has a negligible effect on the results.

\begin{table}[tb]
\centering
\caption{Optimized selection criteria for the signal decay mode \LbJpXiK. The first two requirements 
are applied using the momenta before the corresponding mass constraint. \label{tab:sel}}
\cmsTable{\begin{tabular}[c]{ll}
    Variable                                    & Selection         \\ \hline
    $\abs{m(\prot\pim)-m_{\mathrm{PDG}}(\La)}$  & $<$8\MeV          \\
    $\abs{m(\La\pim)-m_{\mathrm{PDG}}(\Xim)}$   & $<$6\MeV          \\
    \ptlb                                       & $>$11.5\GeV       \\
    \ptjp                                       & $>$6.5\GeV        \\
    \ptxim                                      & $>$2.6\GeV        \\
    \ptla                                       & $>$2.2\GeV        \\
    \ptk                                        & $>$1.2\GeV        \\
    $\mumu\Xim\Kp$ vertex fit probability       & $>$5\%            \\
    $\La\pim$ vertex fit probability            & $>$5\%            \\
    $\prot\pim$ vertex fit probability          & $>$1\%            \\
    \Xim vertex displacement from \Lb vertex    & $>$3\unit{s.d.}   \\
    \La vertex displacement from \Xim vertex    & $>$0\unit{s.d.}   \\
    \Lb vertex displacement from PV             & $>$3\unit{s.d.}   \\
   Angle between \Xim momentum and displacement & $<$0.0447\unit{rad}\\
   Angle between \La momentum and displacement  & $<$0.14\unit{rad} \\
   Angle between \Lb momentum and displacement  & $<$0.0447\unit{rad}\\
   PV impact parameter for pion from \Xim decay & $>$0.4\unit{s.d.} \\
   PV impact parameter for kaon                 & $>$0.4\unit{s.d.} 
\end{tabular}}
\end{table}

\section{Invariant mass distributions}

\begin{figure}[b]
\centering
    \includegraphics[width=0.48\textwidth]{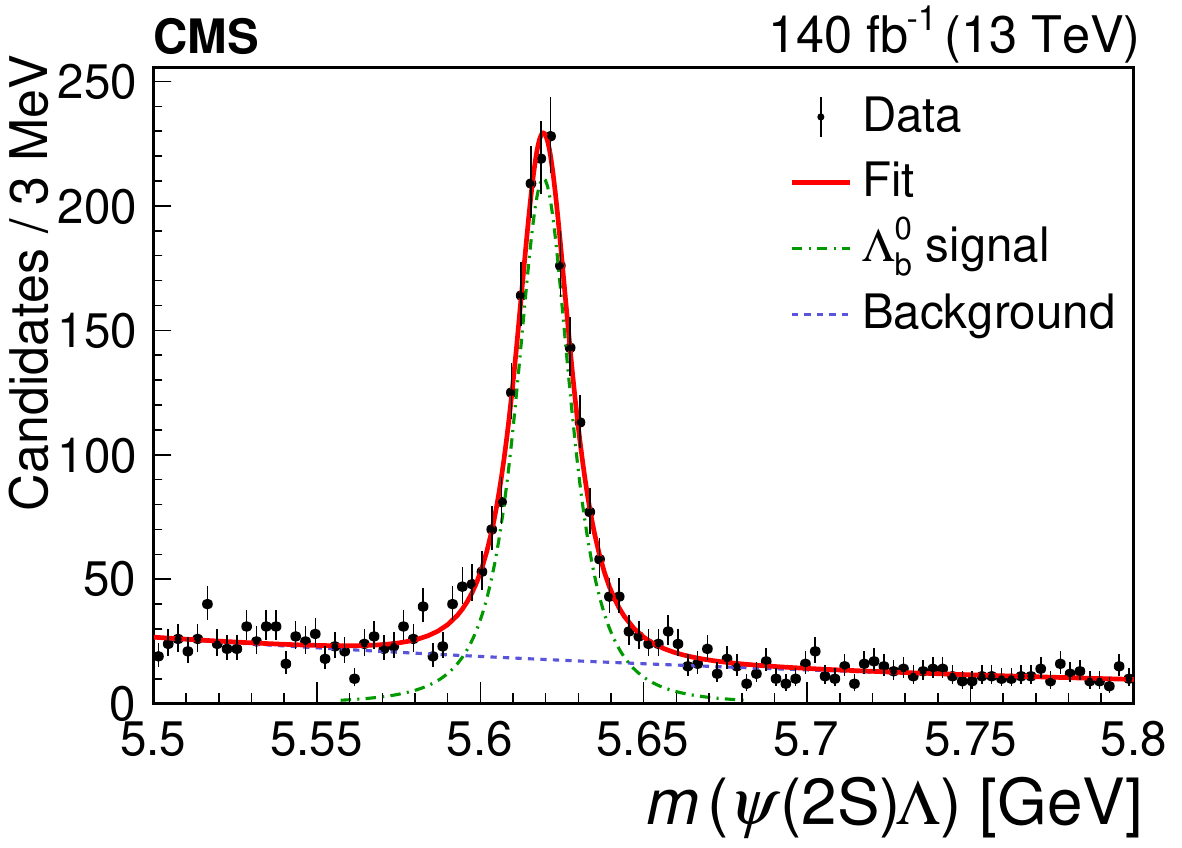}
    \includegraphics[width=0.48\textwidth]{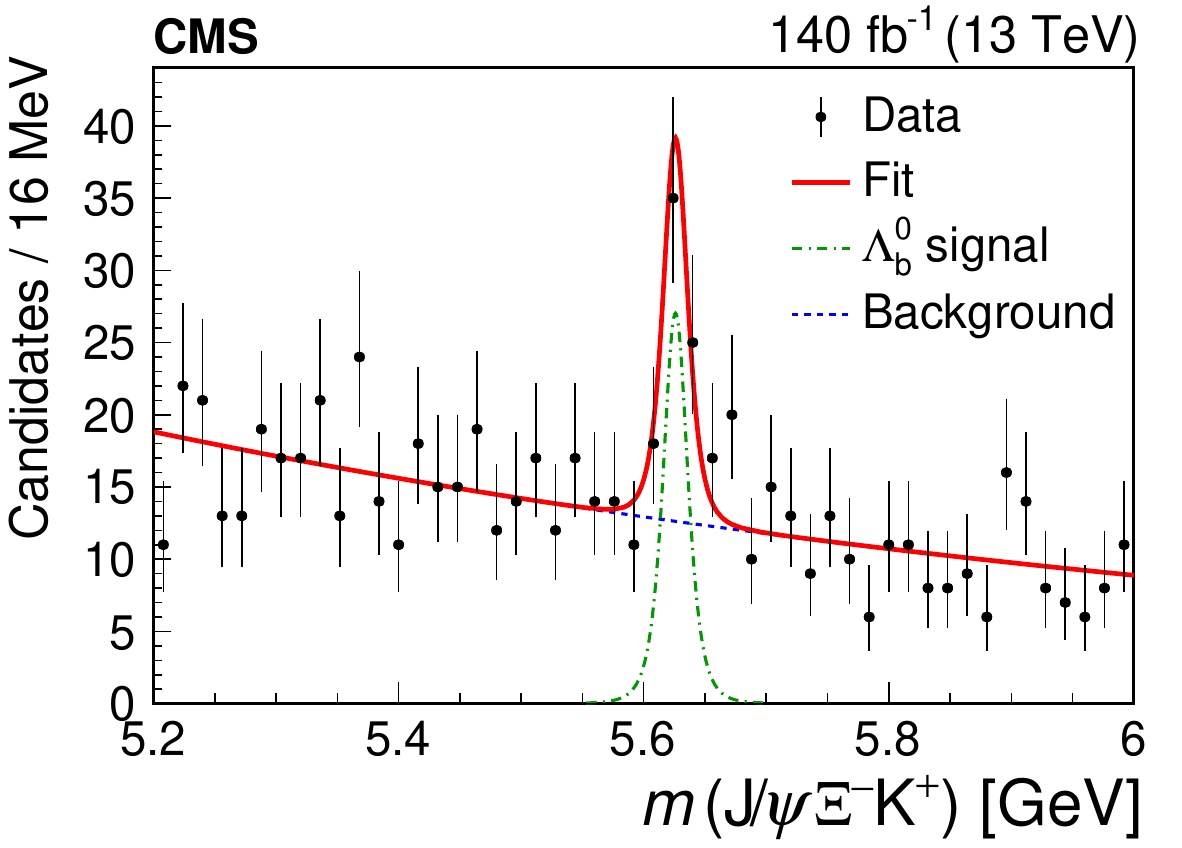}
    \caption{Measured \PsiLam (\cmsLeft) and \JpXiK (\cmsRight) invariant mass distributions and
    overlaid fit results.
\label{fig:mainfits}}
\end{figure}

{\tolerance=800
The measured mass distribution of the \PsiLam candidates is shown in Fig.~\ref{fig:mainfits} (left) 
together with the results of an unbinned maximum likelihood fit. 
The signal is modeled with a Student's \textit{t}-distribution~\cite{Studentt} with all parameters (mean, $\sigma$, $n$) free. 
The combinatorial background is described by an exponential function with a free slope parameter and normalization. 
The fitted mass of $5619.3\pm0.3\mev$ is in agreement with the world-average \Lb mass of $5619.60\pm 0.17\mev$~\cite{PDG}, 
and the mass resolution of $8.90\pm0.40\mev$ is slightly larger than, yet in agreement with, 
its value of 8.52\mev found in simulation. The measured yield is $N(\LbPsiLam)=1744\pm63$. 
The \chisq between the binned distribution and the fit function is 76.6 for 94 degrees of freedom, 
demonstrating the good quality of the fit.
\par}

The measured invariant mass distribution of the selected \JpXiK candidates is shown in Fig.~\ref{fig:mainfits} (\cmsRight). 
A narrow peak at the \Lb mass is seen on top of a smooth background.
The \Lb signal is modeled with a Student's \textit{t}-distribution with mean and $\sigma$ floating, 
but the $n$ parameter fixed to the value found by fitting the simulated distribution, 
because of the limited signal yield of $N(\LbJpXiK)=46\pm11$. 
The background is fitted with an exponential function. The \Lb mass returned by the fit 
($5625.9 \pm 3.2\mev$) is within 2\unit{s.d.} of the world-average value~\cite{PDG}.
The width of the signal peak ($\sigma$) is found to be $10.4\pm3.3\mev$, consistent within 1.2\unit{s.d.} 
with the value found in simulation, $6.6\pm0.2\mev$.
The fit quality is good, as demonstrated by the $\chisq/\mathrm{dof}=30.1/45$ for the binned distribution.

The signal significance is evaluated using the likelihood
ratio technique by applying the back\-ground-only and signal-plus-background hypotheses.
In these two fits, a Gaussian constraint is applied on the background shape parameter to the 
one obtained from a fit to the wrong-sign data. 
Similarly, a Gaussian constraint is applied to the signal shape parameter 
$n$ (from simulation) and the resolution $\sigma=\sigma_{\mathrm{MC}}\, (8.90/8.52)$. 
The correction factor is extracted from the normalization channel and 
accounts for the difference in the widths of the peak between the measured and simulated event samples.
The mean value of the peak is also Gaussian-constrained with a central value and 
uncertainty equal to the known \Lb mass and its uncertainty~\cite{PDG}, respectively.
The fit with the signal-plus-background model with these constraints returns a signal yield of $36\pm8$
and is presented in Appendix~\ref{app}. 
Since the conditions of Wilks' theorem~\cite{Wilks:1938dza} are satisfied, 
the asymptotic formulae of Ref.~\cite{Cowan:2010js} 
(Eqs.~(12)~and~(52)) are used to determine the \LbJpXiK signal significance, which is found to be 5.8 standard deviations.
To evaluate the effect of the choice of the model for fitting the signal significance, several alternative 
models of signal and background were tested, including double-Gaussian 
or Johnson~\cite{Johnson} functions for the signal and a second-degree polynomial 
or a modified threshold function for the background. An alternative without 
a constraint on the background shape was also tested. 
The significance obtained with the alternative models varies 
in the range from 5.3 to 5.9 standard deviations. 
This allows us to claim the first observation of the \LbJpXiK decay.

\begin{figure*}[th!]
\centering
    \includegraphics[width=0.32\textwidth]{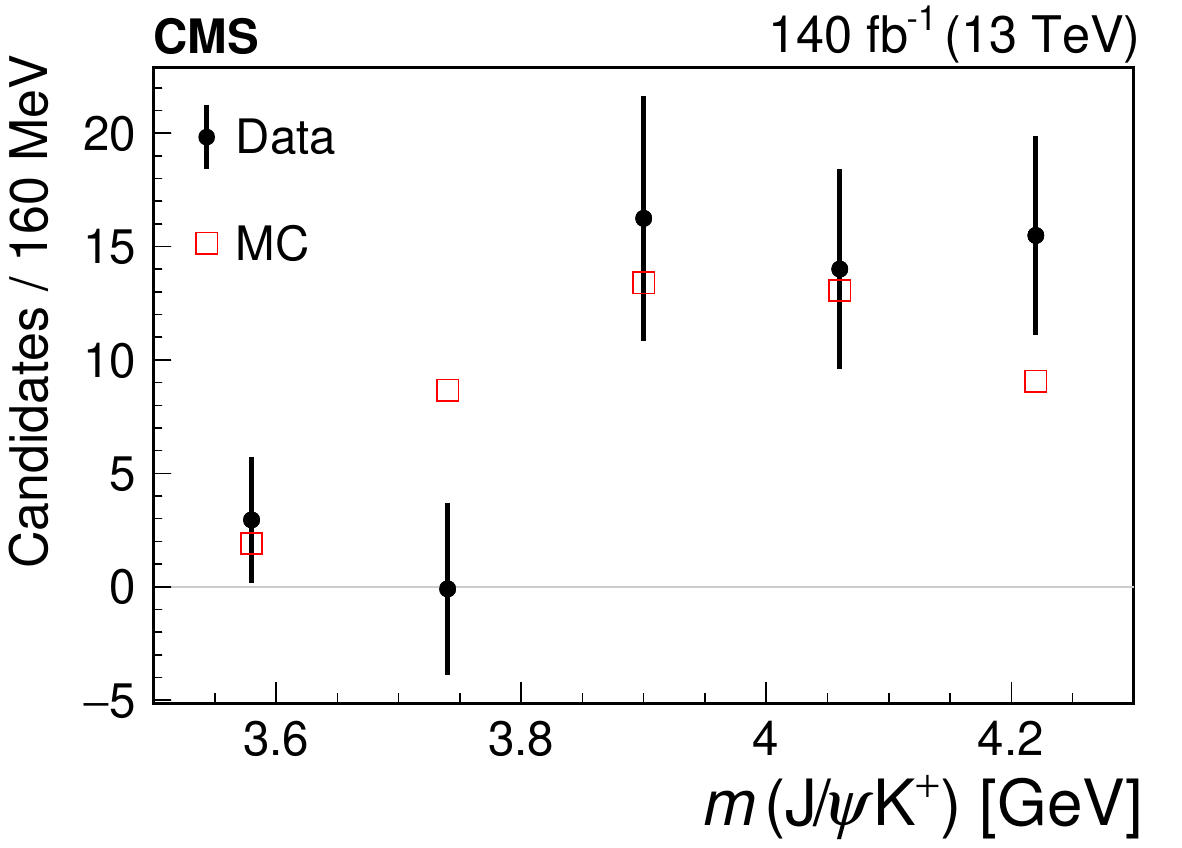}
    \includegraphics[width=0.32\textwidth]{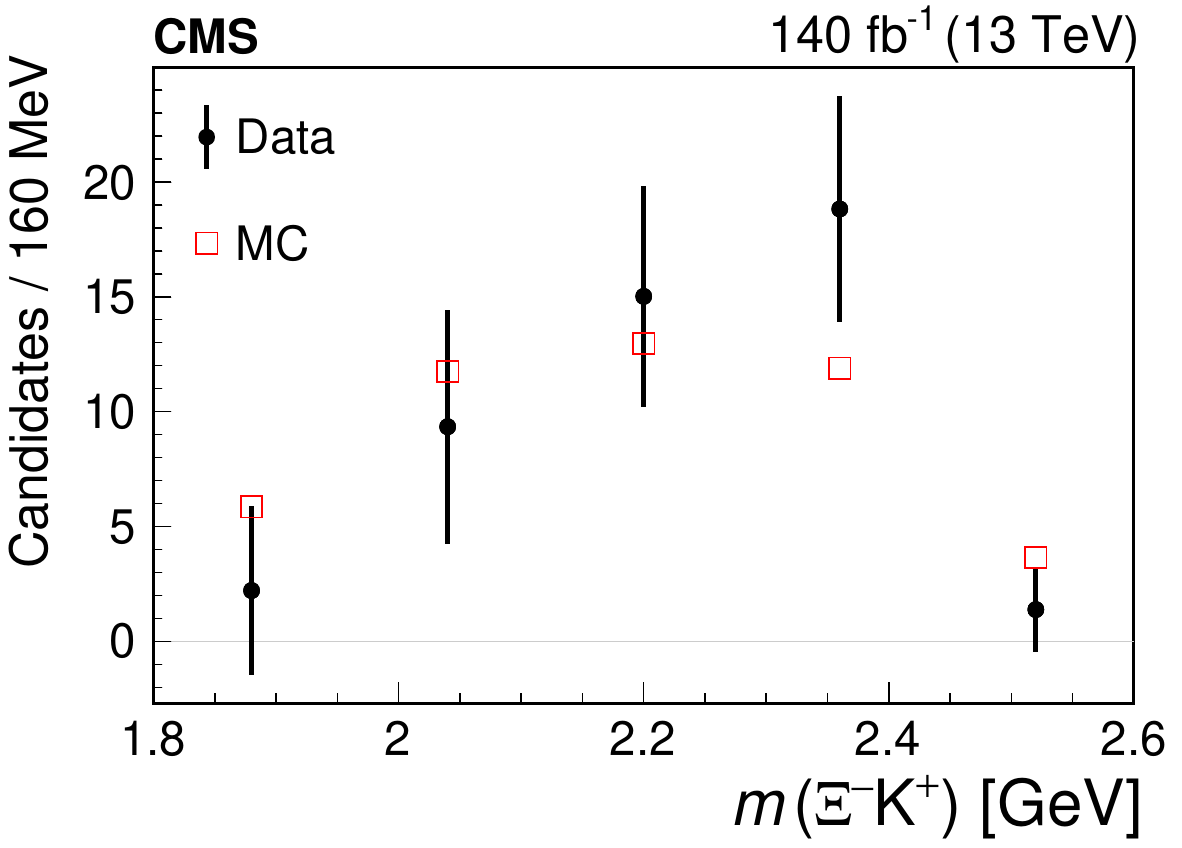}
    \includegraphics[width=0.32\textwidth]{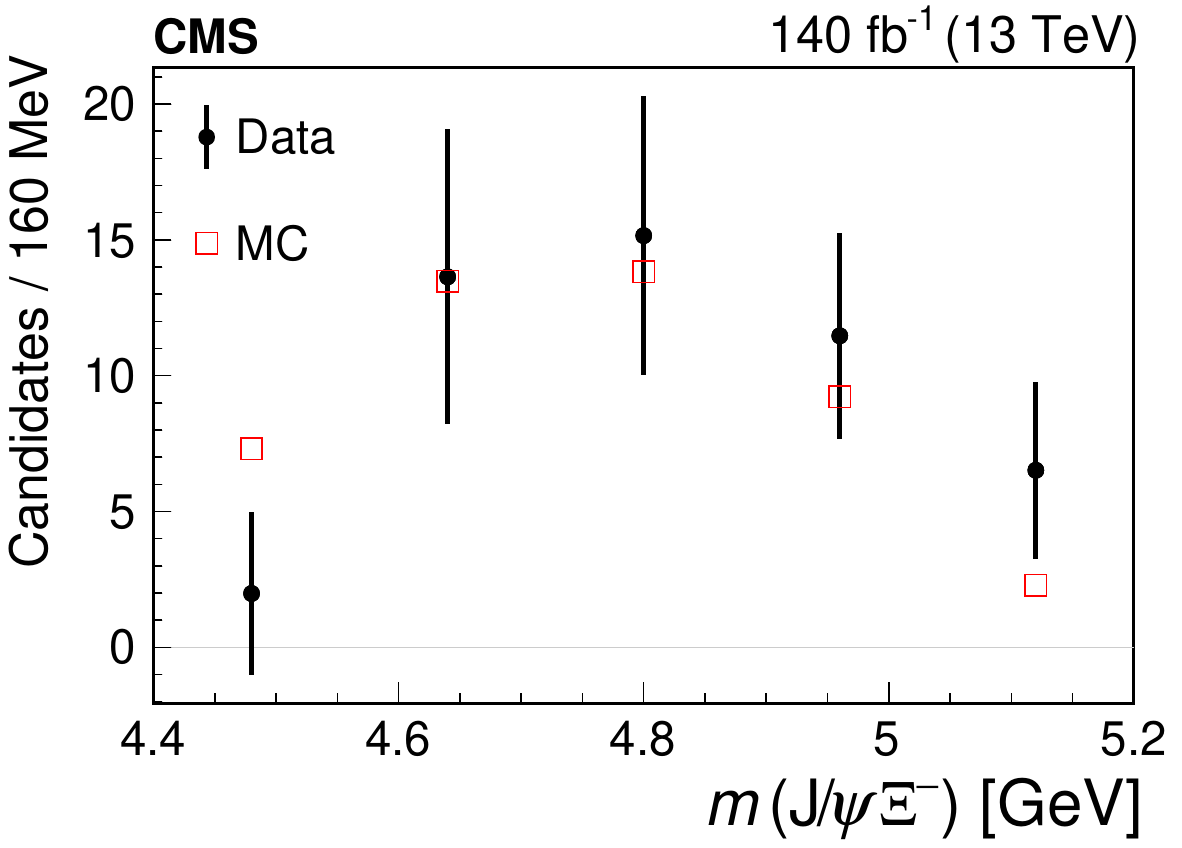}
    \caption{Intermediate invariant mass distributions of the \LbJpXiK decay. 
    The filled circles and empty squares show the measured background-subtracted distributions and the 
    results from the simulation with a phase-space model, respectively.
    \label{fig:splot}}
\end{figure*}

The sensitivity of this analysis to 
potential pentaquark signals in the intermediate invariant mass 
distributions of the \LbJpXiK decay is limited by the low signal yield. 
The background-subtracted two-body invariant mass distributions, obtained with the
\splot technique~\cite{Pivk:2004ty}, are shown in Fig.~\ref{fig:splot}. 
The distributions do not show any clear peaks and agree, within uncertainties, 
with the predictions from the phase space simulation.
The distributions are also consistent with the results of extracting the yields by 
fitting the \Lb signal in each of the five intermediate invariant mass bins.

\begin{figure}[t!]
\centering
    \includegraphics[width=0.48\textwidth]{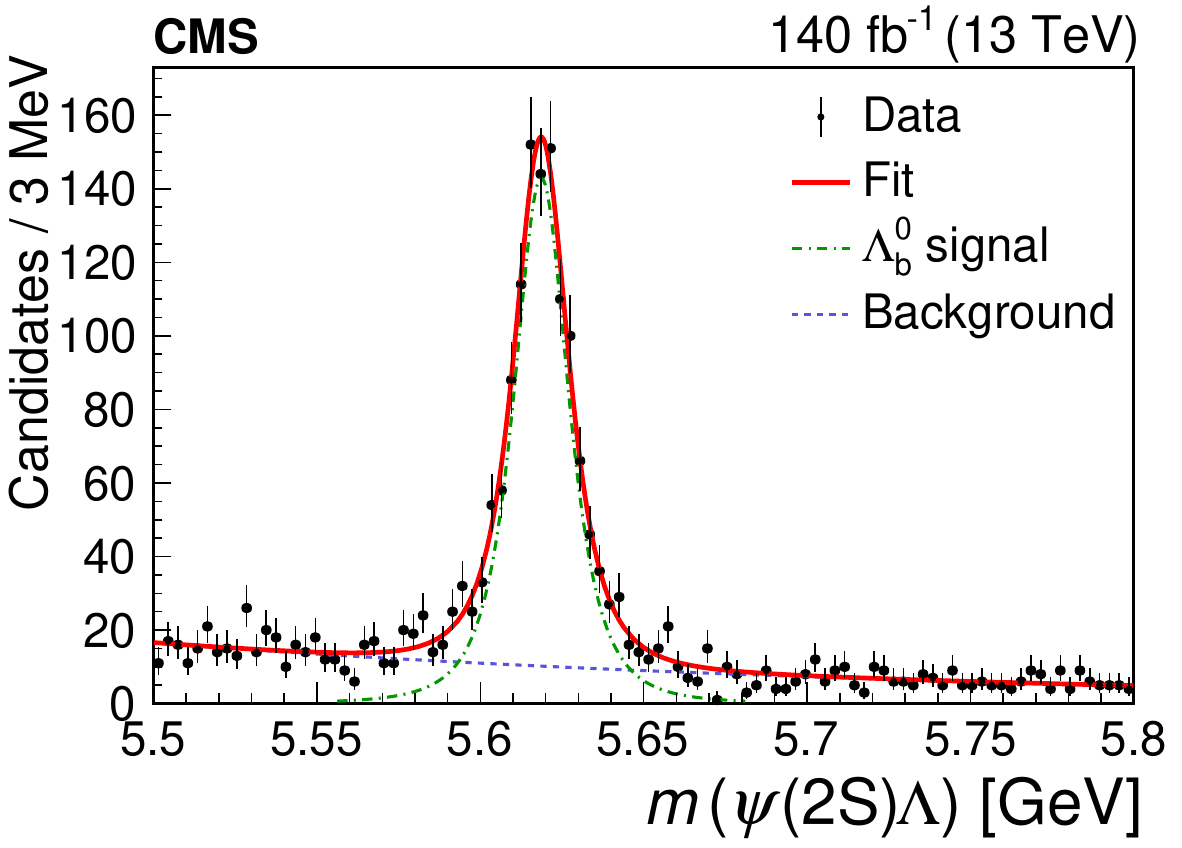}
    \includegraphics[width=0.48\textwidth]{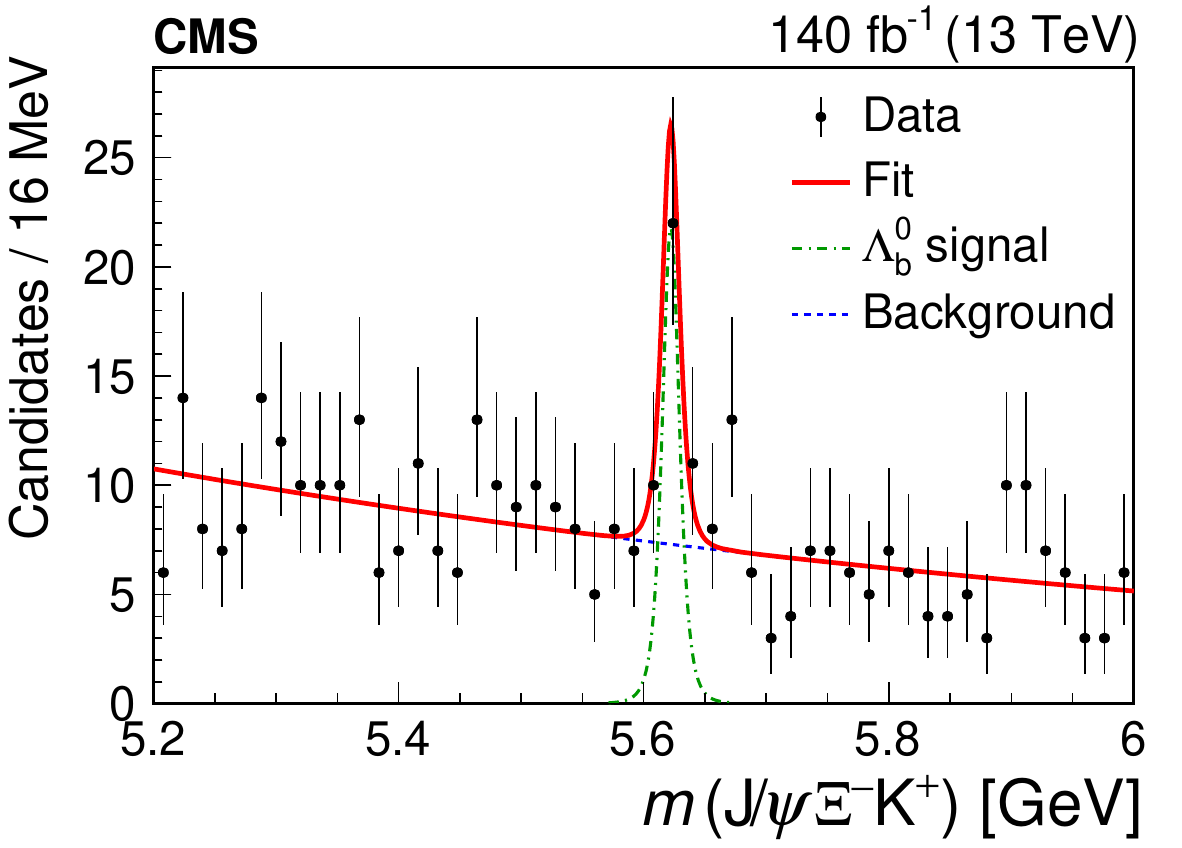}
    \caption{Measured \PsiLam (\cmsLeft) and \JpXiK (\cmsRight) invariant mass distributions and corresponding fits used for the measurement of \rr. 
\label{fig:signalsforR}}
\end{figure}

For the measurement of \rr (Eq.~\eqref{eq:R}), more stringent requirements 
are used, as explained at the end of Section~\ref{sec:evesel}, and the measured signal yields
decrease to $1179\pm47$ and $23\pm7$ for the \LbPsiLam and \LbJpXiK channels, respectively, using unconstrained fits as for Fig~\ref{fig:mainfits}.
These are the baseline results referred to in Section~\ref{sec:syst}. 
The corresponding mass distributions and fits are presented in Fig.~\ref{fig:signalsforR}.

\section{Efficiencies\label{sec:eff}}

Efficiencies for the signal and normalization channels are calculated
using simulated event samples.
The total efficiency is calculated by factorizing into two components: detector acceptance and 
a combined trigger, reconstruction, and selection efficiency.

As only the ratio of the total efficiencies is needed to measure \rr, 
the systematic uncertainties associated with the muon and track
reconstruction are reduced. 
The obtained efficiency ratio is $\epsilon_{\PsiLam}/\epsilon_{\JpXiK} = 5.06\pm0.29$, 
where the uncertainty reflects the limited size of the simulated samples. 
Efficiencies for different years of data-taking are combined with weights
corresponding to the integrated luminosity collected in each year. 
The efficiency for the \LbJpXiK channel is significantly lower than that 
for the \LbPsiLam channel for several reasons including the low energy release in the \XiLaPi decay, 
resulting in a low-momentum pion track. 

\section{Systematic uncertainties \label{sec:syst}}

Many systematic uncertainties, related to the muon reconstruction and identification 
as well as to the trigger efficiency, partially cancel in the measured ratios. 
Since the signal and normalization channels 
have the same number of tracks in the final state, 
most uncertainties related to track reconstruction also cancel 
in the measured ratio \rr. 
However, the \pt spectrum of kaons from the \LbJpXiK decay is observed to differ from 
that of the highest-\pt pion in the \LbPsiLam channel used for normalization. 
Despite the signal and normalization channels having the same number of final-state tracks, 
an uncertainty of 2.3\%~\cite{DP-2022-012} is included, which reflects the difference in 
tracking efficiency between the measured and simulated event samples.
The MC event samples are validated using the normalization channel by comparing the 
measured distributions of variables used in the event selection, after background subtraction,
to those found in simulation; 
no significant discrepancies are found in most of the distributions. 
A small discrepancy was observed in the \ptlb distribution, and the MC event samples for 
both channels were reweighted using \ptlb-dependent weights so that the 
\ptlb distribution in the weighted simulation sample matches the 
background-subtracted distribution measured in the \LbPsiLam channel. 
The efficiency ratio evaluated using these weighted MC samples is found to be 
$\epsilon_{\PsiLam}/\epsilon_{\JpXiK}=4.82\pm0.39$, which is lower by 4.7\% , yet still in 
agreement with the value reported in the previous section. 
An uncertainty of 4.7\% is assigned to account for potential
mismodeling of the \ptlb spectrum. 

The systematic uncertainty related to the choice of the signal model
is evaluated by testing three different models.
For the normalization channel \LbPsiLam the signal shape parameters are floating, 
while for the signal channel \LbJpXiK the mass resolution parameters are fixed to those found in simulation, 
after correcting the width of the peak for the ratio between the two resolutions in the measured and simulated event samples evaluated 
in the normalization channel. 
The tested models simultaneously vary the signal and normalization channels and use a Student's \textit{t}-distribution, a double-Gaussian, and a Johnson function~\cite{Johnson} to model the signal.
The largest deviation in the ratio of the \Lb signal yields from 
the baseline value is taken as the systematic uncertainty.

The systematic uncertainty related to the choice of the background model
is estimated in a similar way, with three alternative models: a second-degree 
polynomial, a threshold function~\cite{CMS:2021rvl, CMS:2020zzv} multiplied by an exponential, 
and a threshold function multiplied by a first-degree polynomial. 

By requiring the \JPP invariant mass to be near the \psit mass, we aim to 
select \LbPsiLam decays. Nevertheless, other $\Lb\to\jpsi\pipi\Lambda$ decays, either from 
different intermediate resonances or from four-body nonresonant decays may contribute. 
To estimate this contribution, we use the \splot technique to subtract the background under 
the \Lb peak and plot the \JPP mass in an expanded mass region. The \JPP mass is 
fitted with a signal component for the \psit and a background component for everything else. 
The integral of the background over the range used to select \psit events yields 30 events, 
which is 2.5\% of the total yield (1179 events). 
This value is used as the systematic uncertainty related to non-\psit contributions in the normalization channel.

The uncertainty in the efficiency ratio due to the limited size of the simulated samples, calculated to 
be 5.6\% in Section~\ref{sec:eff}, is considered as a systematic uncertainty.

In order to assess the reliability of the efficiency evaluation from the simulated samples, 
the selection criteria on muon and \jpsi \pt, dimuon vertex probability, track impact parameter, 
and \pt of the soft pion from \Xim decay are tightened, one at a time, until the 
signal efficiency decreases by 10 or 20\% with respect to that obtained with the selection used for the \rr measurement. 
The analysis is repeated each time, and the value of \rr is 
re-calculated and compared to the baseline \rr value. The differences ($d$) 
between the two values and their uncertainties ($\delta d$), which also account for 
the correlation between the two values, are evaluated. 
The largest value of $\sqrt{\smash[b]{d^2-(\delta d)^2}}$ among the different variations of the selection criteria
is found to be 14.3\% and is used as the systematic uncertainty in the efficiency ratio. 

Table~\ref{tab:syst} summarizes the previously discussed systematic uncertainties in the ratio \rr.
The total uncertainty is calculated as the sum in quadrature of the individual sources.

\begin{table}[h!]
\centering
\caption{The relative systematic uncertainties in the measurement of \rr. \label{tab:syst}}
\begin{tabular}[c]{lc}
    Source                          & Uncertainty (\%) \\ \hline
    Tracking efficiency             & 2.3   \\  
    \ptlb spectrum                  & 4.7   \\  
    Signal model                    & 3.9   \\ 
    Background model                & 6.7   \\ 
    Non-\psit contribution          & 2.5   \\ 
    Limited size of MC samples      & 5.6   \\
    Selection efficiency            & 14.3  \\[\cmsTabSkip]
    Total                           & 18.2  \\[\cmsTabSkip]
\end{tabular}
\end{table}

\section{Branching fraction ratio measurement\label{sec:results}}

{\tolerance=600
The branching fraction of the newly observed \LbJpXiK decay, with respect to the \mbox{\LbPsiLam} one, is
measured using Eq.~\eqref{eq:R} to be
\begin{linenomath}
\begin{equation*}\begin{aligned}
\ifthenelse{\boolean{cms@external}}
{ 
\mathcal{R}  &\equiv \frac{\mathcal{B}(\LbJpXiK)}{\mathcal{B}(\LbPsiLam)}  \\ 
&= [3.38\pm 1.02\stat\pm 0.61\syst\pm 0.03\,\ebran]\%,
} 
{ 
\mathcal{R}  \equiv \frac{\mathcal{B}(\LbJpXiK)}{\mathcal{B}(\LbPsiLam)} = 
[3.38\pm 1.02\stat\pm 0.61\syst\pm 0.03\,\ebran]\%,
} 
\end{aligned}\end{equation*}
\end{linenomath}
where the last uncertainty is related to the uncertainties in the branching fractions 
$\mathcal{B}(\PJPP)=34.68\pm0.30\%$ and $\mathcal{B}(\XiLaPi)=99.887\pm 0.035\%$~\cite{PDG}. 
\par}

\section{Summary}

{\tolerance=800
The \LbJpXiK decay is observed with a significance exceeding 5 standard 
deviations using $\sqrt{s}=13\TeV$ proton-proton collision data corresponding to an integrated luminosity 
of $140\fbinv$ collected by the CMS experiment.
The branching fraction is measured with respect to the \LbPsiLam decay to be 
$\mathcal{B}(\LbJpXiK)/\mathcal{B}(\LbPsiLam) = [3.38\pm 1.02\stat\pm 0.61\syst\pm 0.03\,\ebran]\%$.
The distributions of intermediate invariant masses $m(\jpsi\Xim)$, $m(\jpsi\Kp)$, and $m(\Xim\Kp)$ 
from the \LbJpXiK decay are also presented.
This is the first discovered multibody decay containing the $\jpsi\Xim$ system, which 
opens the possibility to search for doubly-strange hidden-charm pentaquarks when more data are collected.
The new results are important for understanding the strong interaction 
processes in hadronic decays of beauty baryons and the possible formation of exotic multiquark states.
\par}

\begin{acknowledgments}
We congratulate our colleagues in the CERN accelerator departments for the excellent performance of the LHC and thank the technical and administrative staffs at CERN and at other CMS institutes for their contributions to the success of the CMS effort. In addition, we gratefully acknowledge the computing centers and personnel of the Worldwide LHC Computing Grid and other centers for delivering so effectively the computing infrastructure essential to our analyses. Finally, we acknowledge the enduring support for the construction and operation of the LHC, the CMS detector, and the supporting computing infrastructure provided by the following funding agencies: SC (Armenia), BMBWF and FWF (Austria); FNRS and FWO (Belgium); CNPq, CAPES, FAPERJ, FAPERGS, and FAPESP (Brazil); MES and BNSF (Bulgaria); CERN; CAS, MoST, and NSFC (China); MINCIENCIAS (Colombia); MSES and CSF (Croatia); RIF (Cyprus); SENESCYT (Ecuador); ERC PRG, RVTT3 and TK202 (Estonia); Academy of Finland, MEC, and HIP (Finland); CEA and CNRS/IN2P3 (France); SRNSF (Georgia); BMBF, DFG, and HGF (Germany); GSRI (Greece); NKFIH (Hungary); DAE and DST (India); IPM (Iran); SFI (Ireland); INFN (Italy); MSIP and NRF (Republic of Korea); MES (Latvia); LAS (Lithuania); MOE and UM (Malaysia); BUAP, CINVESTAV, CONACYT, LNS, SEP, and UASLP-FAI (Mexico); MOS (Montenegro); MBIE (New Zealand); PAEC (Pakistan); MES and NSC (Poland); FCT (Portugal); MESTD (Serbia); MCIN/AEI and PCTI (Spain); MOSTR (Sri Lanka); Swiss Funding Agencies (Switzerland); MST (Taipei); MHESI and NSTDA (Thailand); TUBITAK and TENMAK (Turkey); NASU (Ukraine); STFC (United Kingdom); DOE and NSF (USA).
    
\hyphenation{Rachada-pisek} Individuals have received support from the Marie-Curie programme and the European Research Council and Horizon 2020 Grant, contract Nos.\ 675440, 724704, 752730, 758316, 765710, 824093, and COST Action CA16108 (European Union); the Leventis Foundation; the Alfred P.\ Sloan Foundation; the Alexander von Humboldt Foundation; the Science Committee, project no. 22rl-037 (Armenia); the Belgian Federal Science Policy Office; the Fonds pour la Formation \`a la Recherche dans l'Industrie et dans l'Agriculture (FRIA-Belgium); the Agentschap voor Innovatie door Wetenschap en Technologie (IWT-Belgium); the F.R.S.-FNRS and FWO (Belgium) under the ``Excellence of Science -- EOS" -- be.h project n.\ 30820817; the Beijing Municipal Science \& Technology Commission, No. Z191100007219010 and Fundamental Research Funds for the Central Universities (China); the Ministry of Education, Youth and Sports (MEYS) of the Czech Republic; the Shota Rustaveli National Science Foundation, grant FR-22-985 (Georgia); the Deutsche Forschungsgemeinschaft (DFG), under Germany's Excellence Strategy -- EXC 2121 ``Quantum Universe" -- 390833306, and under project number 400140256 - GRK2497; the Hellenic Foundation for Research and Innovation (HFRI), Project Number 2288 (Greece); the Hungarian Academy of Sciences, the New National Excellence Program - \'UNKP, the NKFIH research grants K 124845, K 124850, K 128713, K 128786, K 129058, K 131991, K 133046, K 138136, K 143460, K 143477, 2020-2.2.1-ED-2021-00181, and TKP2021-NKTA-64 (Hungary); the Council of Science and Industrial Research, India; ICSC -- National Research Center for High Performance Computing, Big Data and Quantum Computing, funded by the EU NexGeneration program (Italy); the Latvian Council of Science; the Ministry of Education and Science, project no. 2022/WK/14, and the National Science Center, contracts Opus 2021/41/B/ST2/01369 and 2021/43/B/ST2/01552 (Poland); the Funda\c{c}\~ao para a Ci\^encia e a Tecnologia, grant CEECIND/01334/2018 (Portugal); the National Priorities Research Program by Qatar National Research Fund; MCIN/AEI/10.13039/501100011033, ERDF ``a way of making Europe", and the Programa Estatal de Fomento de la Investigaci{\'o}n Cient{\'i}fica y T{\'e}cnica de Excelencia Mar\'{\i}a de Maeztu, grant MDM-2017-0765 and Programa Severo Ochoa del Principado de Asturias (Spain); the Chulalongkorn Academic into Its 2nd Century Project Advancement Project, and the National Science, Research and Innovation Fund via the Program Management Unit for Human Resources \& Institutional Development, Research and Innovation, grant B37G660013 (Thailand); the Kavli Foundation; the Nvidia Corporation; the SuperMicro Corporation; the Welch Foundation, contract C-1845; and the Weston Havens Foundation (USA).    
\end{acknowledgments}

\bibliography{auto_generated} 

\appendix

\section{Invariant mass distribution fit with constraints\label{app}}
\begin{figure}[hb]
\centering
    \includegraphics[width=0.48\textwidth]{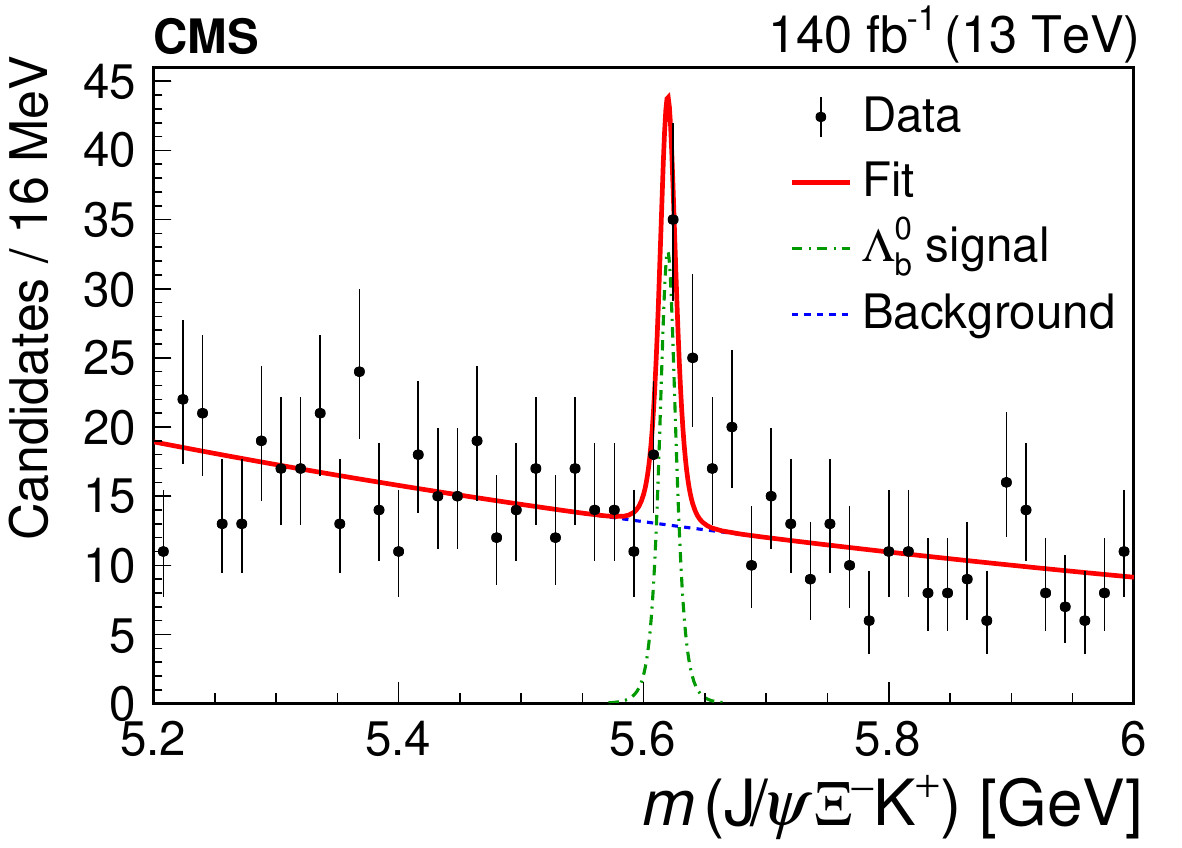}
    \caption{Measured \JpXiK invariant mass distribution and overlaid constrained fit result.
\label{fig:app}}
\end{figure}

The fit of the measured \JpXiK invariant mass distribution 
with constraints on the background shape parameter, signal shape parameter,
resolution, and the mean value of the peak is presented in Fig.~\ref{fig:app}.
The fit quality is good, as demonstrated by $\chisq/\mathrm{dof}=35.6/44$ for the binned distribution.

\cleardoublepage \section{The CMS Collaboration \label{app:collab}}\begin{sloppypar}\hyphenpenalty=5000\widowpenalty=500\clubpenalty=5000
\cmsinstitute{Yerevan Physics Institute, Yerevan, Armenia}
{\tolerance=6000
A.~Hayrapetyan, A.~Tumasyan\cmsAuthorMark{1}\cmsorcid{0009-0000-0684-6742}
\par}
\cmsinstitute{Institut f\"{u}r Hochenergiephysik, Vienna, Austria}
{\tolerance=6000
W.~Adam\cmsorcid{0000-0001-9099-4341}, J.W.~Andrejkovic, T.~Bergauer\cmsorcid{0000-0002-5786-0293}, S.~Chatterjee\cmsorcid{0000-0003-2660-0349}, K.~Damanakis\cmsorcid{0000-0001-5389-2872}, M.~Dragicevic\cmsorcid{0000-0003-1967-6783}, A.~Escalante~Del~Valle\cmsorcid{0000-0002-9702-6359}, P.S.~Hussain\cmsorcid{0000-0002-4825-5278}, M.~Jeitler\cmsAuthorMark{2}\cmsorcid{0000-0002-5141-9560}, N.~Krammer\cmsorcid{0000-0002-0548-0985}, D.~Liko\cmsorcid{0000-0002-3380-473X}, I.~Mikulec\cmsorcid{0000-0003-0385-2746}, J.~Schieck\cmsAuthorMark{2}\cmsorcid{0000-0002-1058-8093}, R.~Sch\"{o}fbeck\cmsorcid{0000-0002-2332-8784}, D.~Schwarz\cmsorcid{0000-0002-3821-7331}, M.~Sonawane\cmsorcid{0000-0003-0510-7010}, S.~Templ\cmsorcid{0000-0003-3137-5692}, W.~Waltenberger\cmsorcid{0000-0002-6215-7228}, C.-E.~Wulz\cmsAuthorMark{2}\cmsorcid{0000-0001-9226-5812}
% ...
\par}
\vskip\cmsinstskip
\dag:~Deceased\\
$^{1}$Also at Yerevan State University, Yerevan, Armenia\\
$^{2}$Also at TU Wien, Vienna, Austria\\
% ...
\end{sloppypar}
\end{document}